\documentclass[twocolumn,prb,a4paper]{revtex4}
\usepackage{amsmath}
\usepackage{graphics}

\begin{document}
\title{What can be learned from the schematic mode-coupling approach
\\to experimental data ?}
\author{V. Krakoviack}
\altaffiliation[Present address: ]{CUC\textsuperscript{3}, 
Department of Chemistry, University of Cambridge, Lensfield Road, 
Cambridge CB2 1EW, United Kingdom}
\author{C. Alba-Simionesco}
\affiliation{Laboratoire de Chimie Physique, b{\^a}t. 349,
universit\'{e} de Paris-Sud, 91405 Orsay cedex, France}

\begin{abstract}
We propose a detailed investigation of the schematic mode-coupling
approach to experimental data, a method based on the use of simple
mode-coupling equations to analyze the dynamics of supercooled
liquids. Our aim here is to clarify different aspects of this approach
that appeared so far uncontrolled or arbitrary, and to validate the
results obtained from previous works. Analyzing the theoretical
foundations of the approach, we first identify the parameters of the
theory playing a key role and obtain simple requirements to be met by
a schematic model for its use in this context. Then we compare the
results obtained from the schematic analysis of a given set of
experimental data with a variety of models and show that they are all
perfectly consistent. A number of potential biases in the method are
identified and ruled out by the choice of appropriate models. Finally,
reference spectra computed from the mode-coupling theory for a model
simple liquid are analyzed along the same lines as experimental data,
allowing us to show that, despite the strong simplification in the
description of the dynamics it involves, the method is free from
spurious artifacts and provides accurate estimates of important
parameters of the theory. The only exception is the exponent
parameter, the evaluation of which is hindered, as for other methods,
by corrections to the asymptotic laws of the theory present when the
dynamics is known only in a limited time or frequency range.
\end{abstract}
\maketitle

\section{Introduction}
Since the mode-coupling theory (MCT) of the liquid-glass transition
has been proposed in the mid-eighties \cite{leu84pra,bengotsjo84jpc},
considerable work has been done in order to compare its predictions
with experimental data and computer simulation results
\cite{leshouchesgotsjo92rppgot99jpcm}.

Most of these tests of the MCT are based on the use as fitting
functions of the quasi-universal laws that are obtained from an
asymptotic analysis of the family of non-linear equations to which the
mode-coupling equations belong and that are valid only in the vicinity
of the predicted so-called ideal glass transition singularity. For
experimental data, this is essentially the only practicable approach,
since typical glass-formers are molecular liquids for which a
first-principle study is out of reach and since most experiments, like
dynamic light-scattering spectroscopy
\cite{cumlidupicdre96prelebdreguipiccum97zpb}, probe complex
mechanisms in which various types of dynamical processes are
entangled. In the case of computer simulations, the situation appears
more comfortable, since simple systems can be considered for which all
the data needed for an extensive comparison with the theory can be
directly obtained with high precision. However, it turns out that one
still has to rely on asymptotic results of the theory at some point.
Indeed, one finds that the MCT is not able to predict accurately the
location of the putative ideal glass transition: it has thus to be
determined from the simulation results, usually with a fit to an
asymptotic law, in order to calibrate the comparison between theory
and simulation \cite{naukob97pre}.

Such a use of the asymptotic laws does not go without difficulties,
originating in the fact that the ideal glass transition is never
observed as such: it is usually argued to be avoided because of the
existence of so-called ``activated processes'' not taken into account
by the theory, whose effect is to restore the ergodicity of the system
in the predicted ideal glassy state. Thus the transition is expected
to be replaced by a smooth crossover between two dynamical regimes,
one dominated by the mode-coupling mechanism and the other dominated
by activated processes. A first problem with the occurrence of these
additional processes is that it hinders a clear separation of time
scales between the fast transient dynamics and the structural
relaxation, with which MCT exclusively deals, and thus makes the
domain where the asymptotic regime of MCT is expected more difficult
to resolve. In addition, the study of the corrections to the
asymptotic laws shows that these laws are valid only in a very small
domain around the ideal glass transition
\cite{frafucgotmaysin97pre,fucgotmay98pre}: it can thus be expected
that the dynamics in this domain will be markedly affected by
crossover effects. Finally, from a more pragmatic point of view, since
the asymptotic laws of the theory apply to rather specific aspects of
the dynamics (for instance, in the $\alpha$ relaxation regime, since
the shape of the relaxation itself is non-universal, the main
quantitative asymptotic prediction deals with the variation of the
relaxation time with temperature), only selected parts of the
available experimental information are usually involved in most
experimental tests of the MCT on molecular glassformers, a situation
which can be considered quite unsatisfactory. For these reasons, it
appears desirable to have a means to test the MCT on experimental data
which is not limited to the use of the critical predictions of the
theory and which is able to give an overall account of the dynamics.
This can precisely be achieved within the so-called schematic approach
to experimental data.

Schematic mode-coupling models are simple sets of mode-coupling
equations retaining the essential non-linear structure of the
equations of the MCT in one or two equations and with only a few
parameters. They played a major role in the development of the theory
to investigate the universal properties of the general mode-coupling
equations and to demonstrate the potentialities of the theory
\cite{dergot86jpc,gotsjo87jpc,gothau88zpb,gotsjo89ajpcm,gotsjo89bjpcm}.
Recently, they have been found useful as quantitative models for
numerical fits of experimental data covering a wide time or frequency
range \cite{albkra95jcp,albkrakramigrom96jrs,kraalbkra97jcp,%
fragotmaysin97pre,sinligotfucfracum98jncs,rufbeatouecolesmar98jncs,%
tei96prl,gotvoi00pre,note1}. The basic idea behind this schematic
approach to experimental data is that, since all equations of the
mode-coupling type share distinctive universal properties, meaningful
informations are likely to be obtained if a suitable minimal set of
equations of this family is used as an effective mode-coupling model
to describe experimental data.  One interest of the method is then
that the universal properties of the mode-coupling equations are used
in a milder sense than in their usual direct applications: it is thus
possible to numerically take into account corrections to the
asymptotic laws that can be important if the system is too far from
its putative ideal glass transition. In addition, one can tentatively
describe the full dynamics, including the non-universal $\alpha$
relaxation and the fast transient dynamics (although through rather
crude approximations), and thus possibly overcome the problem of
insufficient separation of time scales. For these reasons, the
schematic approach appears as a method of choice to critically
investigate the validity of the previous tests of the theory based on
the asymptotic laws.

Of course, the power of the method comes at a price: even a minimal
schematic model (in a sense to be defined latter) will have numerous
adjustable parameters. This makes the approach numerically extremely
flexible, and one finds for instance that it is possible to give an
account of the experimental dynamics with an ideal MCT schematic model
even at low temperatures where ideal MCT is known inadequate. Thus one
should not overestimate the physical meaning of the fitted
parameters. However, as was found in previous works, by critically
looking at the results of a schematic analysis, one can address from a
renewed point of view the problem of testing the predictions of the
MCT on experimental data.

In the present paper, the third of a series devoted to the schematic
approach to experimental data \cite{albkra95jcp,kraalbkra97jcp}, we
aim to address various issues that have been overlooked in previous
works, in order to better settle and understand the principles and
results of this approach. In the next section, we give an overview of
the schematic studies proposed so far and of the models on which they
were based. In Sec.~\ref{sec3}, a set of simple requirements that need
to be fulfilled in order to define a suitable minimal schematic model
is proposed. In Sec.~\ref{sec4}, we examine the influence of the
choice of the schematic model on the results of the analysis of a
given set of experimental data, and Sec.~\ref{sec5} is devoted to a
tentative validation of the method, by analyzing model spectra with
well defined properties within the schematic approach. We finally
present our conclusions and draw perspectives for the approach.

\section{Models for schematic studies}
\label{sec2}

In this section, we give an overview of the schematic studies proposed
so far. This gives us the opportunity to settle a few definitions and
notations to be used latter.

The first schematic study has been proposed by Al\-ba-Simionesco and
Krauzman \cite{albkra95jcp}, where they analyzed low-frequency Raman
scattering spectra of the fragile glassformer metatoluidine with
various two-correlator schematic models. One of them, based on the
so-called F$_{12}$ model \cite{got84zpb,note2}, is defined by the
equations
\begin{widetext}
\begin{subequations}\label{f12model}
\begin{gather}
\ddot\phi_i(t)+\Gamma_i\Omega_i\dot\phi_i(t)+\Omega_i^2
\phi_i(t)+\Omega_i^2\int^t_0 m_i(t-\tau)\dot\phi_i(\tau)
d\tau=0,\quad i=0,1\label{genlan}\\
m_0(t)=v_1\phi_0(t)+v_2\phi_0(t)^2,\label{m0}\\
m_1(t)=rm_0(t),\label{m1}
\end{gather}
\end{subequations}
\end{widetext}
with initial conditions $\phi_i(0)=1$, $\dot\phi_i(0)=0$, $i=0,1$, and
has subsequently been used in studies of Raman spectra of other
glassformers \cite{albkrakramigrom96jrs} and, more importantly, of
depolarized light-scattering spectra of salol and CKN
[$\mbox{Ca}_{0.4}\mbox{K}_{0.6}(\mbox{NO}_3)_{1.4}$] covering a wide
frequency range (more than four decades) \cite{kraalbkra97jcp}. In all
these works, where other choices for the memory-function $m_0$ have
occasionally been considered \cite{albkra95jcp}, the fitting function
was defined, up to an amplitude factor, as the susceptibility
associated with the effective correlator
\begin{equation}\label{phischem}
\phi_{\text{sch}}(t)=\gamma\phi_0(t)^2+(1-\gamma)\phi_1(t)^2.
\end{equation}

Another widely used model differs only from the preceding one through
the choice
\begin{equation}
m_1(t)=v_s\phi_0(t)\phi_1(t),
\end{equation}
an expression first proposed by Sj{\"o}gren \cite{sjo86pra}, and the use
as the fitting function of the susceptibility associated to $\phi_1$,
with an adjustable amplitude factor. It has been used to study
depolarized light-scattering spectra of glycerol
\cite{fragotmaysin97pre} and orthoterphenyle (OTP)
\cite{sinligotfucfracum98jncs}, as well as coherent neutron-scattering
and Brillouin spectra of $\mathrm{Na_{0.5}Li_{0.5}PO_3}$
\cite{rufbeatouecolesmar98jncs}. An extended MCT version
of this model has also been used recently to study depolarized
light-scattering, dielectric-loss and incoherent neutron-scattering
spectra of propylene carbonate \cite{gotvoi00pre}.

Finally, self-intermediate scattering functions obtained by molecular
dynamics simulations of a nickel-zirconium model have been analyzed by
Teichler \cite{tei96prl} with a single-correlator model with $m_0$
defined as
\begin{equation}
m_0(t)=\gamma\left(e^{\mu\phi_0(t)}-1\right).
\end{equation}
At variance with the works mentioned above, a retarded damping term
involving four adjustable parameters has been used in the present case
instead of the simple Markovian approximation made in
Eq.~\eqref{genlan} involving the single parameter $\Gamma_0$.

Looking at the equations, one sees that these models involve \emph{a
priori} a large number of fitting parameters. To reduce this number,
simplifying assumptions are usually made: in
Refs.~\onlinecite{albkra95jcp}, \onlinecite{albkrakramigrom96jrs},
\onlinecite{kraalbkra97jcp}, and \onlinecite{fragotmaysin97pre}, the
only temperature-dependent parameters are chosen to be $v_1$ and
$v_2$, whereas in Ref.~\onlinecite{sinligotfucfracum98jncs}, these are
$v_2$ and $v_s$, and in Ref.~\onlinecite{rufbeatouecolesmar98jncs},
$v_1$, $v_2$ and $v_s$ change with temperature, the latter depending
also on the wave vector modulus.

\section{Definition of a suitable minimal schematic model}
\label{sec3}

In all schematic studies proposed so far, a certain degree of
arbitrariness has often been involved in the choice of the details of
the model to be used. The most serious case relates perhaps to the use
of the memory function \eqref{m1} in our preceding works
\cite{albkra95jcp,albkrakramigrom96jrs,kraalbkra97jcp}, which, at
variance with Sj{\"o}gren's model\cite{sjo86pra}, cannot be given any
precise theoretical foundations. But other minor choices can also be
questioned. There is indeed no physical reason to prefer using a
simple correlator as the fitting function instead of a composite one
like in Eq.~\eqref{phischem}, the apparent advantage of the former
choice, which seems to make the relation between the fitting function
and the dynamical variable probed by the experiment transparent,
relying on a mere tautology. One can also wonder why, for similar
experimental data and schematic equations, the effective vertex
entering the second memory function ($r$ or $v_s$) has been held
constant with temperature in Refs.~\onlinecite{kraalbkra97jcp} and
\onlinecite{fragotmaysin97pre} and not in
Ref.~\onlinecite{sinligotfucfracum98jncs} for instance, where it is
$v_1$ which has been kept fixed. In most cases, these various choices
have been only validated \emph{a posteriori} by the success of the
model to fit the data under study and/or the obtention of parameters
whose variations match the expectations from the microscopic theory
\cite{rufbeatouecolesmar98jncs,gotvoi00pre}.

There are nevertheless a few difficulties that could be expected,
related to the possible existence of non-generic features of the
chosen model. Indeed, as stated in the introduction, generic
mode-coupling equations display remarkable universal properties
\cite{leshouchesgotsjo92rppgot99jpcm}, but, when a schematic model is
considered, because it is very simple, one can experience severe
limitations in the realization of these universal features: certain
values of important parameters of the theory can be unattainable or
\emph{ad hoc} correlations between these parameters can be imposed. An
example of the first situation is provided by the F$_n$ models, $n>2$,
for which $1/2$ is the only possible value of the exponent parameter
$\lambda$, whereas it can take any value in the interval $[1/2,1]$ for
the general mode-coupling equations. The second situation is realized
with the F$_{12}$ model, where the exponent parameter and the critical
non-ergodicity parameter $f_0^c$ are always related through
$f_0^c=1-\lambda$.

For these reasons, it seems desirable to understand what are the
minimal requirements that must be met by a schematic model in order to
avoid such biases and possible consequent inconsistencies in the
results of its application. We propose here a series of criteria based
on a discussion of the asymptotic properties of the mode-coupling
equations in the vicinity of the ideal glass transition. For this, we
consider the case of an experiment probing the evolution of a
normalized correlation function $\phi_A$ associated to an unspecified
dynamical process in a given glassforming liquid, when a thermodynamic
parameter $X$ (temperature or density for instance) is varied. Note
that the following discussion can easily be extended to situations
where more than one observable are considered in a unified framework
\cite{rufbeatouecolesmar98jncs,gotvoi00pre}, with exactly the same
conclusions.

By looking at the asymptotic predictions of the the\-ory
\cite{leshouchesgotsjo92rppgot99jpcm}, a few important parameters
emerge. On the one hand, one finds parameters that characterize the
liquid under study and the thermodynamic path along which the
experiment is performed. These are $X_c$, the value of $X$ at the
ideal glass transition, $C_X$, the proportionality constant relating
the separation parameter $\sigma$ defined by the theory to $X$ through
the relation
\begin{equation}
\sigma\simeq C_X\frac{X-X_c}{X_c},
\end{equation}
and $\lambda$, the exponent parameter governing numerous aspects of
the critical dynamics. On the other hand, one finds parameters that
are specific to the correlation function $\phi_A$: $f^c_A$, the
non-ergodicity parameter at the transition, and $h_A$, the so-called
critical amplitude. Thus we identify five parameters that are expected
to be of special relevance. The first three of them are involved in
the description of the relative evolution of the two characteristic
time scales of the dynamics in the ergodic phase, $t_\sigma$ and
$\tau_\sigma$ characterizing the $\beta$ regime and the $\alpha$
relaxation respectively. This can be seen from the ratio
\begin{equation}\label{ratio}
\frac{\tau_\sigma}{t_\sigma}=B^{-1/b} |C_X|^{-1/(2b)}
\left|\frac{X-X_c}{X_c}\right|^{-1/(2b)},
\end{equation}
where $b$ and $B$ are uniquely determined by $\lambda$
\cite{got90jpcm}.  In addition, $\lambda$ alone determines the shape
of the dynamics in the $\beta$ regime. The two remaining parameters,
$f^c_A$ and $h_A$, are related to the amplitudes of the
$\alpha$ relaxation and of the $\beta$ process, as shown for instance
by the scaling relation
\begin{equation}
\phi_A(t)=f^c_A+h_A |C_X|^{1/2} \left|\frac{X-X_c}{X_c}\right|^{1/2}
g_\lambda(t/t_\sigma),
\end{equation}
describing the $\beta$ process and the onset of $\alpha$ relaxation in
the ergodic phase close to the singularity.

But in fact, this parameter counting is only valid when the full
mode-coupling equations describing the dynamics of the system are
known. Two differences appear when the only available information is
the function $\phi_A$ at different $X$ values. First, it can be seen
that $h_A$ never appears alone in the asymptotic laws, but always
inside products $h_A|C_X|^{1/2}$. For this reason, the relevant
parameter is rather the physical amplitude
$H_A=h_A\left|C_X/(1-\lambda)\right|^{1/2}$, which appears for
instance in the relation between $X$ and the non-ergodicity parameter
in the ideal glassy state,
\begin{equation}
f_A=f^c_A+H_A \left|\frac{X-X_c}{X_c}\right|^{1/2}.
\end{equation}
This remark would be irrelevant if it was possible to know $C_X$ from
the data at hand, but this is not the case. Indeed, according to the
theory, the shape of the $\alpha$ relaxation is non-universal. This
means that there exists an undetermined proportionality constant
relating $\tau_\sigma$, the value emerging from the theoretical
calculation, to the phenomenological $\alpha$ relaxation time
$\tau_\alpha(X)$, which can be defined for instance after
$\phi_A(\tau_\alpha)=1/e$ and which is the only type of value that can
be accessed from the knowledge of $\phi_A$ alone. Such a problem does
not occur for the $\beta$ process which is universal, i.e. it depends
only on $\lambda$. This means that the ratio \eqref{ratio} can only be
known up to an undetermined factor and thus that the value of $C_X$
cannot be obtained. This is in fact an important result, since it
shows that one can know the absolute distance (in terms of the
separation parameter $\sigma$) of a system to its ideal glass
transition only if the mode-coupling equations describing the full
dynamics are known.

Moving now back to the schematic approach to experimental data, one
finds thus that, in order to analyze the evolution of the correlation
function $\phi_A$ with an effective schematic correlator
$\phi_{\text{sch}}$, the model must be able to reproduce the values of
$\lambda$, $f^c_A$ and $H_A$ characterizing the system and the
observable under investigation. A point on the transition surface of
the schematic model must exist where one will have
\begin{equation}
\lambda_{\text{sch}}=\lambda, \ f^c_{\!\text{sch}}=f^c_A, \ 
H_{\text{sch}}=h_{\text{sch}}\left(\frac{|C_{X,\text{sch}}|}
{1-\lambda_{\text{sch}}}\right)^{1/2}=H_A.
\end{equation}
The last equality is not a strong constraint since $h_{\text{sch}}$
and $C_{X,\text{sch}}$ relate to decoupled aspects of the
description of the dynamics, the first one being defined at the
transition point, whereas the second one describes the approach of
this point. One can thus expect that the interplay of these two
aspects will allow to compensate possible non-generic features
relative to $h_{\text{sch}}$ (like in the case of the F$_{12}$ model
where $h_{\text{sch}}=\lambda_{\text{sch}}$). The first two
constraints are on the contrary very strong and need to be taken into
account. One has in particular to worry about the accessible domain
for the values of $\lambda_{\text{sch}}$ and $f^c_{\text{sch}}$ and to
avoid the introduction of arbitrary correlations between these
parameters.

Eventually, one finds thus that a schematic model needs to have at
least three adjustable coupling constants, at least one of them
changing with temperature. Indeed, with a single coupling constant,
only one pair ($\lambda_{\text{sch}}$,$f^c_{\text{sch}}$) is
possible. With two coupling constants, a relation linking
$\lambda_{\text{sch}}$ and $f^c_{\text{sch}}$ exists, and for a given
value of $\lambda_{\text{sch}}$, only a finite number of
$f^c_{\text{sch}}$ values will exist. One needs three coupling
parameters to be able to decouple $\lambda_{\text{sch}}$ and
$f^c_{\text{sch}}$.

We can now look at the schematic models introduced in the previous
section at the light of these new results. It is easy to see that
these conditions are matched by the model used in
Ref.~\onlinecite{kraalbkra97jcp} and defined by Eqs.~\eqref{f12model}
and \eqref{phischem}, since it relies on four coupling parameters
($v_1$, $v_2$, $r$ and $\gamma$) to obtain the values of
$\lambda_{\text{sch}}$, $f^c_{\text{sch}}$ and $H_{\text{sch}}$. This
is not the case of the model used by Teichler \cite{tei96prl}, which
has only two coupling parameters $\gamma$ and $\mu$, and one indeed
finds inconsistencies in the results: for instance, the exponent
parameter $\lambda$ appears to change with the considered
observable. It seems thus likely that the success of the calculation
is essentially to be credited to the flexibility of the fitting
procedure (in particular through the choice of a four-parameter
retarded friction) and to the consideration of observables with high
plateau values which probably turn out to be compatible with the value
of the real exponent parameter.  About the other studies, one finds in
fact that the present discussion is inoperant, because only part of
the dynamics is considered: the $\alpha$ relaxation is not present in
Refs.~\onlinecite{albkra95jcp} and \onlinecite{albkrakramigrom96jrs},
and the fast transient dynamics is not included in
Refs.~\onlinecite{fragotmaysin97pre},
\onlinecite{sinligotfucfracum98jncs}, and
\onlinecite{rufbeatouecolesmar98jncs}, with the effect of releasing in
both cases the constraint on the value of $f^c_{\text{sch}}$.

An interesting consequence of these constraints is that they restrain
the possibility to give a precise physical meaning to the separate
components of a schematic model. This is especially clear when a
composite correlator like Eq.~\eqref{phischem} is used. Indeed, the
relations between the components of the fitting function are expected
to be mainly imposed by the need to reproduce the values of $\lambda$,
$f^c_A$ and $H_A$: it is then hardly conceivable that the value of the
coupling constant $\gamma$ or the effect of the details of the
expression for $\phi_{\text{sch}}$ will reflect any physical reality,
for instance in terms of the light scattering mechanism. This is of
course less obvious when the fitting function is a pure correlator,
because of the tautological relation between the fitting function and
the dynamical variable probed by the experiment, but, for the previous
reasons, we believe that in general a schematic model has to be
considered in its globality as a minimal phenomenological model
allowing to give an account of the dynamics of a system in terms of
mode-coupling effects, and nothing more has to be expected from it.

To conclude this part, it is worth stressing that the present
discussion refers only to the asymptotic predictions of the theory and
that problems originating in pre-asymptotic corrections or the
existence of activated processes have been ignored. Moreover, only a
few numbers have been considered. In particular, the precise shape of
the relaxation (particularly in the $\alpha$ relaxation domain) has
not been taken into account, whereas it can make an \emph{a priori}
well-suited model totally inefficient for the schematic approach.  For
these reasons, the preceding remarks are only guidelines aimed at
rationalizing and optimizing the use of schematic models, but they
offer no guarantee of success.

\section{Schematic analysis of experimental spectra with different models}
\label{sec4}

In the present section, we propose a study of depolarized
light-scattering spectra of a fragile glassformer by means of a
variety of schematic models. The aim of this work is twofold. First,
in all previous schematic analysis, only one model has ever been used
at a time for a given set of experiments and no systematic
investigation of the influence of the choice of the model on the
results has been proposed so far. Secondly, in the previous section, a
few parameters have been isolated that are expected to play a special
role in the description of the dynamics of a given observable. The use
of different schematic models to analyze the same set of experimental
data, depending of their ability to provide consistent estimates for
these values, should thus allow us to confirm or infirm the
corresponding discussion.

The data we have studied are the CKN depolarized light-scattering
spectra measured by Cummins \emph{et al.} \cite{liduchecumtao92pra}
and already analyzed with a schematic model in
Ref.~\onlinecite{kraalbkra97jcp}. The reason CKN has been chosen (and
not salol for instance) for this study is the low intensity of its
$\alpha$ peak, which is slightly smaller than the microscopic peak:
this guarantees that, when the spectra are studied with a composite
effective correlator like in Eq.~\eqref{phischem}, the amplitudes of
the contributions of both terms will not be too different and thus
this allows one to obtain a better contrast between the outcome of
different schematic analysis. To limit the influence of the avoidance
of the ideal glass transition, only rather high temperatures, between
383 and 468 K, have been considered. Compared to our previous work,
the two lower temperatures (363 and 373 K) have been excluded, leading
to small differences in the results (within the error bars of the
method). Note that at the temperatures we have considered, one can
expect the spectra to be free of the influence of the spurious
experimental artifacts evidenced in
Ref.~\onlinecite{surwienovrossok98prb}.

Only schematic models closely related to the one used in
Ref.~\onlinecite{kraalbkra97jcp} and with the same number of
parameters have been considered. As in this previous work, they have
been used to perform simultaneous fits of the experimental data at the
nine selected temperatures: the whole dynamical scenario developing as
temperature varies is thus encoded in the only temperature dependent
parameters of the schematic models, which are the effective vertices
entering in the expression of the memory function $m_0$.

We first stick to the two-correlator F$_{12}$ model. There are two
routes by which the model can be modified. On the one hand, different
second memory functions can be chosen. We have considered two cases:
\begin{subequations}\label{varmem1}
\begin{align}
  m_1(t)&=r(v_1\phi_0(t)+v_2\phi_0(t)^{2}),\label{varmem1a}\\
  m_1(t)&=v_s\phi_0(t)\phi_1(t).\label{varmem1b}
\end{align}
\end{subequations}
On the other hand, one can vary the expression of the effective
correlator as a function of $\phi_0$ and $\phi_1$. We have considered
four possibilities:
\begin{subequations}\label{varcorreff}
\begin{align}
\phi_{\text{sch}}(t)&=\gamma\phi_0(t)^2+(1-\gamma)\phi_1(t)^2,
\label{varcorreffa}\\
\phi_{\text{sch}}(t)&=\gamma\phi_0(t)^2+(1-\gamma)\phi_1(t),
\label{varcorreffb}\\
\phi_{\text{sch}}(t)&=\gamma\phi_0(t)+(1-\gamma)\phi_1(t)^2,
\label{varcorreffc}\\
\phi_{\text{sch}}(t)&=\gamma\phi_0(t)+(1-\gamma)\phi_1(t).
\label{varcorreffd}
\end{align}
\end{subequations}
In previous works \cite{albkra95jcp,kraalbkra97jcp}, the choice of a
specific expression was originally motivated by assumptions on the
light-scattering mechanism. However, from the discussion of the
preceding section, it seems very unlikely that this question is of any
relevance: indeed, since individually each correlator has no real
physical meaning, the precise way it enters the calculation is clearly
unimportant.

\begin{figure}
\rotatebox{270}{\includegraphics*[12.5cm,1.5cm][19.5cm,10cm]{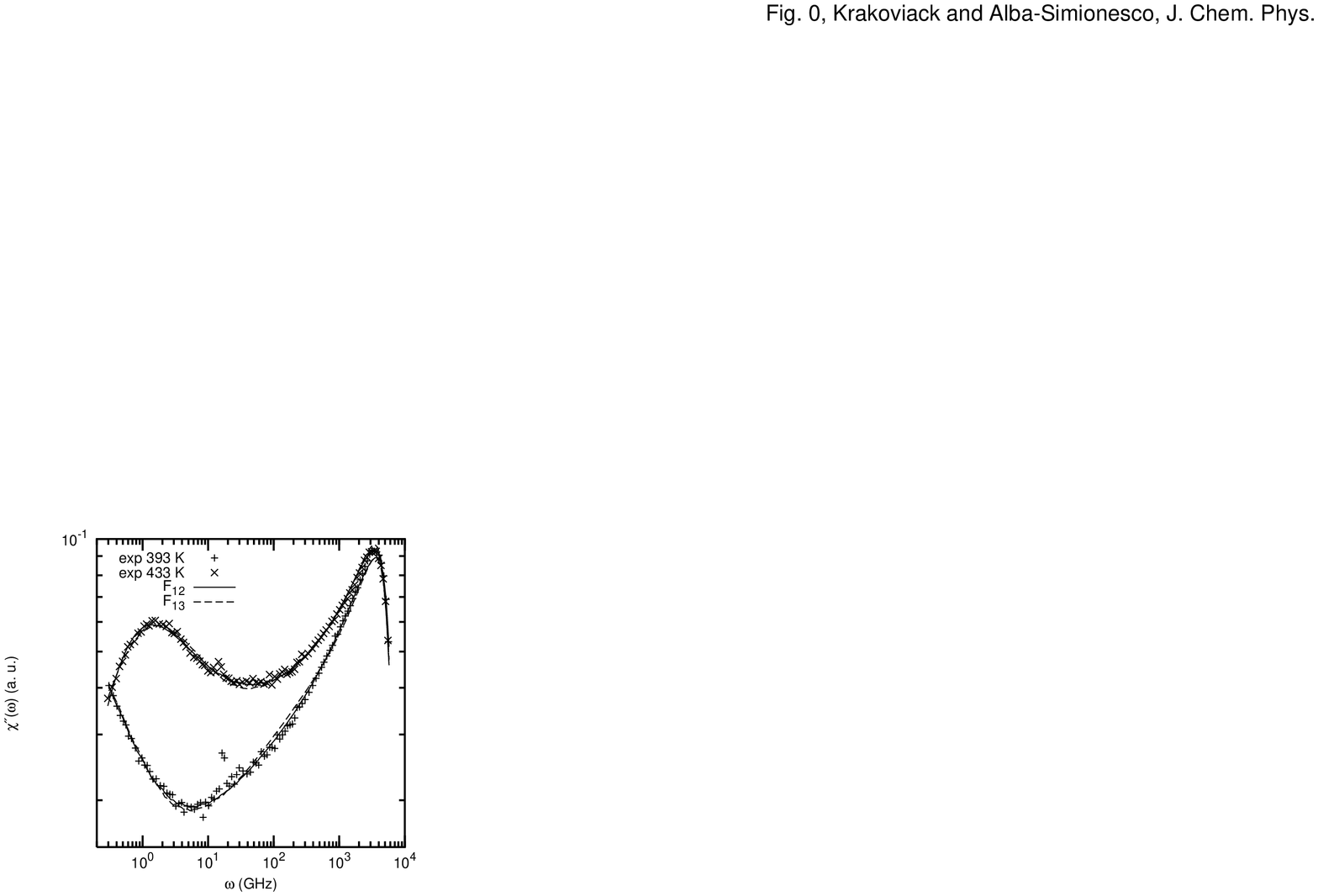}}
\caption{\label{fitexp} Experimental depolarized light-scattering
  spectra of CKN at 393 and 433 K (for readability, the data are
  reported every 4 points) and corresponding fitting curves obtained
  from a global fit of the nine spectra measured between 383 and 468 K
  with two schematic models of type I, based respectively on the
  F$_{12}$ and F$_{13}$ memory functions.}
\end{figure}
\begin{figure*}
\rotatebox{270}{\includegraphics*{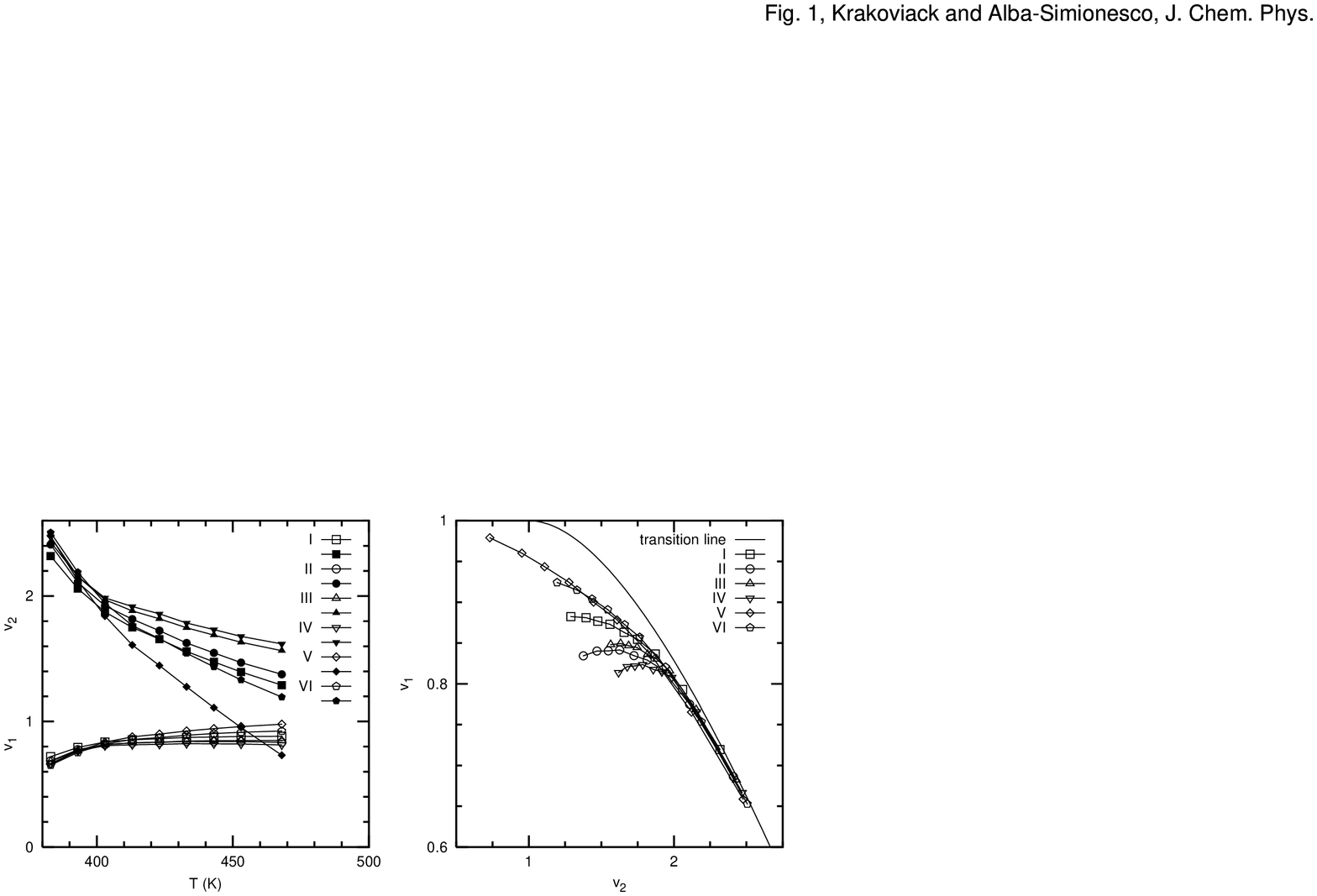}}
\caption{\label{vert1} Effective vertices obtained from a schematic
analysis of the CKN data with various F$_{12}$ models. Left:
temperature evolution of the effective vertices (open symbols: $v_1$,
closed symbols: $v_2$). Right: trajectories in the control parameter
plane $(v_1,v_2)$.}
\end{figure*}
\begin{table}
\bigskip\bigskip
\begin{tabular}{*{7}{c}}
model & I & II & III & IV & V & VI\\ \hline
$m_1$& \multicolumn{4}{c}{\eqref{varmem1a}}&
\multicolumn{2}{c}{\eqref{varmem1b}}\\
\cline{2-5}\cline{6-7}
$\phi_{\text{sch}}$& \eqref{varcorreffa} & \eqref{varcorreffb}&
\eqref{varcorreffc} & \eqref{varcorreffd}& \eqref{varcorreffa} &
\eqref{varcorreffd} \\ \hline $r$ & 8.6 & 3.6
& 5.9 & 2.2 & -- & --\\ $v_s$ & -- & -- & -- & -- & 14.2 & 7.5\\
$\gamma$& 0.30 & 0.29 & 0.31 & 0.22 & 0.32 & 0.49\\\hline
$T_{c}$ (K)   & 388.5  & 388.9   & 388.5& 388.8 & 390.1 &390.8 \\
$\lambda$ & 0.712  & 0.704  & 0.700 & 0.698 & 0.705 &0.701  \\
$f_{\text{sch}}^c$& 0.448  & 0.452  & 0.446  & 0.449 & 0.424 & 0.429 \\
$h_{\text{sch}}$  & 0.779  & 0.696  & 0.887  & 0.799  & 0.657  & 0.707\\
$-C_T$ & 0.150 & 0.183 & 0.113 & 0.133 & 0.230 & 0.150\\ 
$H_{\text{sch}}$ & 0.563 & 0.548 & 0.544 & 0.530 & 0.580 & 0.501
\end{tabular}
\caption{\label{globalt} Temperature-independent coupling constants
and characteristic mode-coupling parameters obtained from a schematic
analysis of the CKN data with various F$_{12}$ models.}
\end{table}

A total of six models (numbered from I to VI, see Table \ref{globalt}
for definition) obtained by various combinations of the previous
expressions have been used. Model I, defined by Eqs.~\eqref{varmem1a}
and \eqref{varcorreffa}, was used in Ref.~\onlinecite{kraalbkra97jcp},
whereas Model VI, using Eqs.~\eqref{varmem1b} and \eqref{varcorreffd},
is close to the one used in Refs.~\onlinecite{fragotmaysin97pre},
\onlinecite{sinligotfucfracum98jncs}, and
\onlinecite{rufbeatouecolesmar98jncs}. The fits are always very good
(see Fig.~\ref{fitexp} for spectra and fits at two representative
sample temperatures) and of equal quality for all models. The
corresponding results for the various coupling constants and relevant
mode-coupling parameters are reported in Table \ref{globalt} and
Fig.~\ref{vert1}.

Each model provides its own series of effective vertices $v_1$ and
$v_2$ and values of the coupling constants $r$ or $v_s$ and $\gamma$,
distinct from those obtained with the other models. The description of
the dynamics in terms of effective mode-coupling parameters is thus
clearly non-unique, as expected from the non-universality of the
$\alpha$ relaxation. For all models, a behavior similar to the one
found in Ref.~\onlinecite{kraalbkra97jcp} is recovered, interpreted as
follows: at the highest temperatures the effective vertices
are linear functions of temperature and show the progressive evolution
of the system towards an ideal glass transition point, but, before
this point is reached, this behavior breaks down, evidencing the
phenomenon of avoidance of the transition and the inadequateness of
the present ideal MCT approach at low temperature. The reader is
referred to previous work for a discussion of the consequences of this
observation \cite{kraalbkra97jcp}.

Looking at the trajectories in the $(v_1,v_2)$ plane, one sees that
the change in behavior of the effective vertices between the high and
low temperature domains is less visible when Sj{\"o}gren's model,
Eq.~\eqref{varmem1b}, is used: in this case, the trajectories seem to
behave rather regularly in the full temperature domain and to bend
gently to follow the transition line, at variance with what is found
from the use of Eq.~\eqref{varmem1a}, where one can easily separate a
domain where $v_1$ does not vary much from a domain where it decreases
rather rapidly. We attribute this qualitative difference between the
two models to the way the temperature dependence is entered in the
schematic model at the level of the second memory function
$m_1$. Indeed, at a given point $(v_1,v_2)$ of the ergodic domain, one
can qualitatively quantify the strength of the mode-coupling term in
Eq.~\eqref{genlan}, $i=1$, by considering the value of $m_1$ for
representative values $f_0$ and $f_1$ of the correlation functions
$\phi_0$ and $\phi_1$, respectively \cite{note3}. When
Eq.~\eqref{varmem1b} is used with constant $v_s$, the corresponding
factor $v_s f_0 f_1$ depends only \textit{implicitly} on temperature
through the variation of $f_0$ and $f_1$ with the position of the
representative point of the system in the control parameter space
$(v_1,v_2)$, whereas, when Eq.~\eqref{varmem1a} is used with constant
$r$, the factor $r(v_1f_0+v_2f_0{}^2)$ shows an additional
\textit{explicit} temperature dependence, since $m_1$ is defined
directly in terms of the coupling constants $v_1$ and $v_2$, and in
fact this explicit dependence dominates the implicit one. If one now
compares the evolution of these factors for both models when one
explores the $(v_1,v_2)$ plane, one easily finds that, in order to
have similar increases in the strength of the mode-coupling terms, one
has to consider trajectories in the case of Sj{\"o}gren's model that lie
above the corresponding trajectories for Eq.~\eqref{varmem1a} and that
come rather close to the transition line, since in its vicinity only
are found significant increases in $f_0$ and $f_1$, hence the
difference in the results obtained with these two models: the
catastrophic behavior of the vertices due to the avoidance of the
transition is less visible in the case of Sj{\"o}gren's model, because
it is by construction already constrained to follow rather closely the
transition line.

Once this subtle non-generic feature has been recognized, one is able
to rationalize the similarities and differences between the results of
the various schematic analysis proposed so far. Since the
light-scattering spectra of glycerol have been analyzed with the same
parameterization of Sj{\"o}gren's model as the one used here for model V
and VI, it is no surprise that the trajectory found by Franosch
\emph{et al.}  \cite{fragotmaysin97pre} is similar to those obtained
here with these two models. This is the case of the study of
$\mathrm{Na_{0.5}Li_{0.5}PO_3}$ as well, despite the choice of a
temperature-dependent $v_s$: above $T_c$, the numerical fits result in
a slow increase of this parameter when temperature is decreased, which
converts, as one would expect, to a trajectory in the $(v_1,v_2)$
plane that is not different from what is obtained when $v_s$ is taken
constant \cite{rufbeatouecolesmar98jncs}. The case of OTP provides
another illustration of this point
\cite{sinligotfucfracum98jncs}. Indeed, in this work, the relative
position of the vertex trajectory and the transition line is imposed
from the start through the choice of a temperature-independent $v_1$,
a constraint that is compatible with the results obtained above for
models I to IV, and, as could be anticipated from the previous
discussion, this coincides with a rapidly increasing $v_s$ when
temperature is decreased, i.e. a strong explicit
temperature-dependence of the second memory function. The same seems
to hold for the extended mode-coupling analysis of G{\"o}tze and
Voigtmann \cite{gotvoi00pre}, where a neat intersection of the ideal
glass transition line by the trajectory in the parameter space of the
schematic model is associated with values of $v_s$ increasing rapidly
with decreasing temperature.

These observations appear at first sight rather anecdotic. They have
nevertheless at least one important physical implication. Indeed,
comparing the results for glycerol and
$\mathrm{Na_{0.5}Li_{0.5}PO_3}$, two liquids with intermediate
fragility, to those for salol, CKN, and OTP, that are all very
fragile, it is tempting to propose a connection between the fragility
of a glassformer and the shape of the effective vertex trajectory that
describes its dynamics. The previous discussion rules out such a
tentative connection and shows that the analogies and differences
found between the various fluids simply originate in the details of
the implementation of the schematic approach.

We now turn to the quantitative analysis of our results. For each
schematic model, the values of the various parameters characterizing
the ideal glass transition have been obtained by extrapolation of the
linear dependence of the effective vertices with temperature at high
temperature, which we have identified as the signature of the domain
where the ideal MCT is adequate (in the next section, support will be
given to this analysis that differs from the one used in
Refs.~\onlinecite{fragotmaysin97pre} and
\onlinecite{rufbeatouecolesmar98jncs}). They are reported in Table
\ref{globalt}. All six models provide consistent estimates for $T_c$,
$\lambda$, $f_{\text{sch}}^c$ and $H_{\text{sch}}$. This consistency
tends to validate the discussion proposed in the previous section
about the relevant variables to describe a given observable in terms
of mode-coupling parameters. This is clearly a non-trivial
observation, since the values of the different parameters entering the
calculation vary notably from one model to another (for instance,
$C_T$ changes by a factor of two between model III and V).

\begin{figure*}
\rotatebox{270}{\includegraphics*{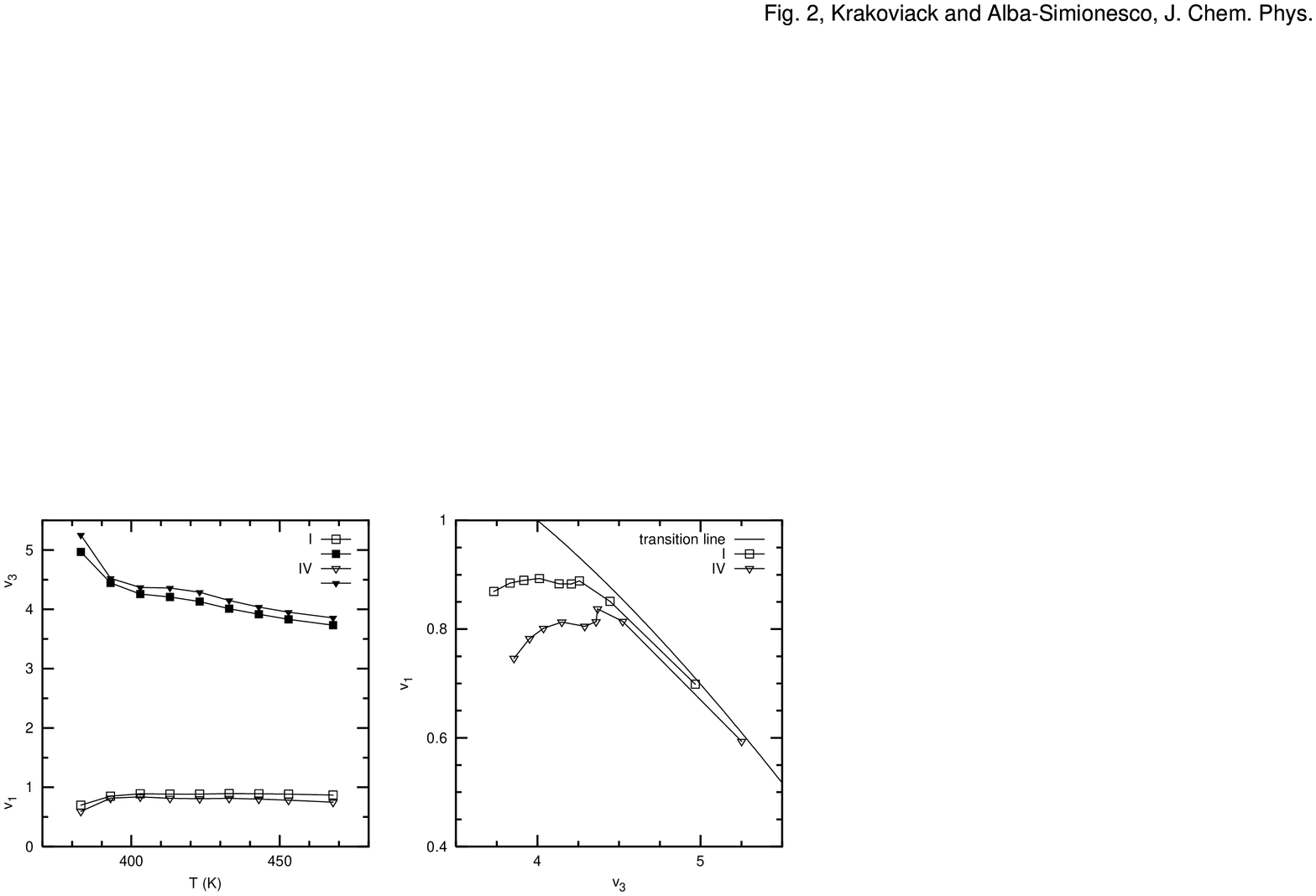}}
\caption{\label{f13fig} Effective vertices obtained from a schematic
analysis of the CKN data with two F$_{13}$ models.  Left: temperature
evolution of the effective vertices (open symbols: $v_1$, closed
symbols: $v_3$). Right: trajectories in the control parameter plane
$(v_1,v_3)$.}
\end{figure*}

In order to investigate stronger changes to the sche\-matic model, we
have considered other memory functions $m_0$, keeping $m_1(t)=rm_0(t)$
as in \eqref{varmem1a}. One of the most studied schematic models after
F$_{12}$ is F$_{13}$ defined as \cite{gotsjo84jpc}
\begin{equation}
m_0(t)=v_1\phi_0(t)+v_3\phi_0(t)^{3}.
\end{equation}
We have studied two models based on this memory function. They will be
denoted I and IV, since they are directly analogous to the preceding I
and IV F$_{12}$ models. We observe a clear deterioration of the
quality of the fits compared to F$_{12}$, but it remains nevertheless
acceptable (see Fig.~\ref{fitexp}). In fact, the use of the F$_{13}$
model is made difficult by one of its non-generic properties. Indeed,
when a B-transition line has a termination point or when two
transition lines meet, strong corrections to the generic asymptotic
laws appear and noticeably affect the shape of the dynamics
\cite{gothau88zpb,gotsjo89bjpcm,fucgothoflat91jpcm}. These corrections
can make it impossible to use a schematic model to fit experimental
data if the description of the system under study requires to come
close to such singularities. In the F$_{12}$ model, a termination
point is found at $v_1=v_2=1$, but here $\lambda=1$, which is quite
far from the values characterizing the typical glassformers, of the
order of 0.7. Thus no problem has to be expected \emph{a priori}. But
in the F$_{13}$ model, one finds that the A- and B-transition lines
meet when $v_1=1$, $v_3=4$, with $\lambda=3/4$. This is rather close
to the values usually associated to experimental systems and one can
thus anticipate difficulties with this model, which we actually see.

\begin{table}
\begin{tabular}{*{3}{c}}
model & I &  IV \\
\hline $r$ & 1.79 & 0.62 \\ $\gamma$& 0.11 & 0.050 \\ \hline
$T_{c}$ (K) & 389.3 & 389.7 \\ $\lambda$ & 0.688  & 0.670 \\
$f_{\text{sch}}^c$& 0.443 & 0.439  \\ $h_{\text{sch}}$ & 0.541  & 0.444
\\ $-C_T$ & 0.254 & 0.373 \\ $H_{\text{sch}}$ &
0.489 & 0.472
\end{tabular}
\caption{\label{f13tab} Temperature-independent coupling constants
and characteristic mode-coupling parameters obtained from a schematic
analysis of the CKN data with two F$_{13}$ models.}
\end{table}

The results we have obtained with the F$_{13}$ models are reported in
Table \ref{f13tab} and in Fig.~\ref{f13fig}. The behavior of the
effective vertices is the same as with F$_{12}$, but slightly less
regular. Critical parameters have been obtained from high temperatures
as before. The agreement with the previous results for the values of
$T_c$ and $f_{\text{sch}}^c$ is very good, but less good for the
values of $\lambda$ and $H_{\text{sch}}$. However one can hardly
conclude from these results, since these parameters characterize the
$\beta$ process, in which the non-generic limitations of the F$_{13}$
model precisely show up.

As it was stressed in the previous section, when using the schematic
approach to experimental data, one has to avoid any arbitrary
correlation between important parameters of the theory introduced in
an \emph{ad hoc} way by the choice of a specific schematic model. In
all models studied so far, one can identify such a correlation in the
fact that, along the B type transition line, an increase of $f_0^c$ is
systematically associated to a decrease of $\lambda$. This could have
introduced a bias in the schematic calculation that would have made
the avoidance of the ideal glass transition line a purely numerical
artifact of the method. Indeed, despite the absence of a true
ergodicity breaking at $T_c$ in real systems, one expects that the
increase of the non-ergodicity parameter below $T_c$ survives under
the form of an increase of the $\alpha$ peak height of the
susceptibility. The vertex drift found with the F$_{12}$ model could
thus be simply traced back to the need to give an account of this
evolution, the obtained trajectories corresponding precisely to an
increase of $f_0^c$ (and thus of $f_{\text{sch}}^c$)
\cite{kraalbkra97jcp}. It would then reflect a misconception and
limitation of the schematic approach itself and not more generally a
failure of the underlying ideal MCT as it was argued in Ref.~15.

To show that our calculations are indeed free of this artifact, we
have investigated a model in which the variations of $\lambda$ and
$f_0^c$ can be decoupled. We chose the F$_{29}$ model, the transition
line of which is made of two branches along which $\lambda$ varies
between $1/2$ and 1: Between the points $v_2^c=4$, $v_9^c=0$ and
$v_2^c=27/7$, $v_9^c=3^9/(7\cdot 2^8)$, $\lambda$ and $f_0^c$ grow
simultaneously, whereas between the points $v_2^c=27/7$,
$v_9^c=3^9/(7\cdot 2^8)$ and $v_2^c=0$, $v_9^c=9^9/8^8$, $\lambda$
decreases and $f_0^c$ increases. Clearly, we have chosen this somewhat
far-fetched model only for its mathematical properties and one could
hardly give any physical motivation for the specific F$_{29}$
functional form. However, we stress that, from the point of view of
the criteria of the preceding section, this model is \emph{a priori}
free from any non-generic features and thus a perfectly acceptable
schematic model: first, it does not show any limitation on the
accessible values of $\lambda$, at variance with all other F$_{pq}$
different from F$_{12}$ ; secondly, in the domain of $\lambda$ we are
interested in (around 0.7), the influence of the higher-order A$_4$
singularity from which its special mathematical features originate is
expected to be negligible.

We have performed schematic analysis with this model in the vicinity
of each branch of the transition line, in an attempt to find if the
coupling of the evolutions of $\lambda$ and $f_0^c$ plays any role in
our calculations. The corresponding fits are quite poor: They only
give a qualitative account of the evolution of the dynamics and it
would seem unreasonable to extract quantitative values for the
critical parameters of the theory from the results. However, a clear
pattern manifests itself: Close to both transition branches, the
behavior of the effective vertices corresponds to a decrease of the
exponent parameter, irrespective of the evolution of $f_0^c$. In fact,
this could have been anticipated from the fact that, below the value
of $T_c$ we have determined, the maximum of the $\alpha$ peak lies
outside the available frequency window, and this eliminates the
requirement to reproduce its amplitude at these low temperatures.
However it is reassuring to show by an independent calculation that
our results are free from this potential bias, that should certainly
be taken into account if data where available on a wider frequency
range.

To summarize, we have shown in this section that different schematic
models applied to a same set of experimental data provide
qualitatively similar results. We have found that they are in good
quantitative agreement for the determination of the values of the
parameters that have been distinguished in the previous section,
provided no non-generic features, the importance of which has been
emphasized for the F$_{13}$ model, interfere with their determination.

\section{Schematic analysis of model spectra}
\label{sec5}

In the previous section, we have tried to validate the schematic
approach and its results showing that different models provide
consistent results when applied to a same set of experimental
data. However, on various aspects, this study is not quite
satisfactory. Indeed, despite favorable indications gained from the
use of the F$_{29}$ model, the meaningfulness of the anomalous
behavior of the effective vertices at low temperature remains to be
assessed. In our previous work \cite{kraalbkra97jcp}, it has been
attributed to the existence of relaxation processes that are not taken
into account in the ideal MCT and that make it impossible to describe
consistently the dynamics of the studied systems with a schematic
model based on this theory. To strengthen this point of view, it would
be necessary to show that the schematic approach is actually able to
give an account of a complex purely mode-coupling dynamics with
smoothly varying fitting parameters, despite of the extreme simplicity
of the schematic equations compared to the complexity of the equations
of the original problem. Moreover, it would be interesting to analyze
data for which the values of the essential parameters $\lambda$,
$f^c_A$ et $H_A$ are known in advance, in order to check to what
extent the schematic approach is able to provide quantitative
estimates for them.

A possibility to perform such a test is to calculate series of curves
from mode-coupling equations known in advance and to analyze them as
one would do for any set of experimental data. It remains then to
compare the results obtained within the schematic approach with the
properties that can be directly determined from the starting
equations. We report the results of such a test in the present
section.

Recently, Franosch \emph{et al.} have proposed a detailed study of the
MCT predictions for the collective dynamics of a simple liquid, with
special emphasis on the asymptotic laws valid in the vicinity of the
ideal glass transition and their corrections
\cite{frafucgotmaysin97pre}. Because their model has been very well
and extensively characterized, it appears clearly as the model of
choice to be the basis of the present evaluation of the schematic
approach. Besides it has already been used in a similar spirit for a
critical study of the method consisting of the use of numerical fits
of the $\beta$ correlator for experimental tests of the MCT
\cite{scitar99jpcm}.

The model is based on the mode-coupling equations for the Brownian
dynamics of the hard-sphere fluid with particle diameter $d$, the
thermodynamic state of which is determined by a single parameter, the
compacity $\eta$, related to the number density $\rho$ through the
relation $\eta=\pi\rho d^3/6$. One starts with the generalized
Langevin equations for the time evolution of the Fourier components of
the normalized density-density correlation functions $\phi_q$,
\begin{subequations} \label{hsmodel}
\begin{equation} \label{hsmodellang}
\tau_q\dot{\phi}_q+\phi_q+\int_0^t d\tau m_q(t-\tau)
\dot{\phi}_q(\tau)=0,
\end{equation}
with the initial condition {\`a} $\phi_q(0)=1$. $\tau_q$ is chosen equal
to $t_{\text{mic}}S_q/(qd)^2$, where $t_{\text{mic}}$ is some
arbitrary molecular time scale and $S_q$ is the structure factor of
the fluid. The mode-coupling memory kernel, defined as a
double integral in Fourier space (thus a six-dimensional integral), is
approximated after moving to bipolar coordinates, by the Riemann sum
of the resulting two-dimensional integral on a grid of $M$ equidistant
values of the wave vector modulus, with a grid step $h$ such that the
$qd$ values vary between $h/2$ and $(2M-1)h/2$. One obtains
\begin{widetext}
\begin{equation}
m_q(t)=\frac{\rho h^3 S_q}{32d^3\pi^2\hat{q}^5}
\sum_{\hat{k}=1}^{M}\ \sum_{\hat{p}=|\hat{q}-\hat{k}|+1}^{\min(M,\
\hat{q}+\hat{k}-1)}\ \hat{k}\hat{p}S_pS_k
\left[(\hat{k}^2+\hat{q}^2-\hat{p}^2)c_k+(\hat{p}^2+\hat{q}^2-
\hat{k}^2)c_p\right]^2 \phi_k(t)\phi_p(t),
\end{equation}
\end{widetext}
\end{subequations}
with $qd=\hat{q}h$, $kd=\hat{k}h$, $pd=\hat{p}h$ and $\hat{q}$,
$\hat{k}$, $\hat{p}\in \{1/2,3/2,\ldots,(2M-1)/2\}$. $c_q$ is the
direct correlation function related to the structure factor through
the Ornstein-Zernike relation $\rho c_q=1-1/S_q$. As in previous
works\cite{frafucgotmaysin97pre}, the static quantities have been
computed within the Percus-Yevick approximation, $M=100$ and $h=0.4$
have been used, and the length and time units have been chosen such
that $d=1$ and $t_{\text{mic}}=160$. One finds then that the ideal
glass transition of the model occurs at a compacity $\eta_c$ comprised
between $0.51591213$ and $0.51591214$.

This model aims to represent the dynamics of a hard-sphere colloidal
suspension and has been shown to compare very favorably at a
quantitative level with the experimental data of van Megen \emph{et
  al} \cite{megund93prl}.

The equations \eqref{hsmodel} have been solved following the algorithm
described in Ref.~\onlinecite{fucgothoflat91jpcm}, then the associated
susceptibilities have been computed by Fourier transform using Filon's
algorithm. We concentrate on the susceptibilities, for which it has
been shown that the range of validity of the asymptotic laws is
usually smaller than for the corresponding correlation functions
\cite{frafucgotmaysin97pre}, in order to be as close as possible to
the situations in which the schematic approach is usually applied to
experimental data.

Five values of the compacity such that the distances to the ideal
glass transition are $\varepsilon_n=(\eta_c-\eta_n)/\eta_c
=10^{-n/3},\ n=4,5,6,7,8$ have been considered. Larger $n$ values have
been excluded, since they would correspond to spectra covering a much
wider frequency range compared to the experimental situation. We have
analyzed spectra for three values of the wave vector modulus:
$q_1=7.0$, corresponding to the main peak of the structure factor,
$q_2=10.2$, corresponding to the first minimum after the main peak,
and $q_3=14.6$, where certain pre-asymptotic corrections seem to be
rather small.

Having now our reference spectra to be used as input like experimental
data would be, each series of curves at a given $q$ value has been
fitted with the two correlator F$_{12}$ model referred to as model IV
in the previous section. For compatibility reasons, no inertial term
is included, i.e. one has
\begin{equation}
\tau_i\dot{\phi}_i+\phi_i+\int_0^t d\tau m_i(t-\tau)
\dot{\phi}_i(\tau)=0,\qquad i=0,1.
\end{equation}
Note that an inertial term would be required both in the microscopic
theory and in a schematic model to study a molecular glassformer, the
dynamics of which is Newtonian, and to reproduce its oscillatory
dynamics in the Raman band or in the boson-peak region. However, for
the goal we wish to achieve here, i.e.~an assessment of the shape of
the vertex trajectory and the calculation of the critical parameters
of the theory from model theoretical spectra, the fact that it has
been consistently omitted in the original model and in the schematic
equations is irrelevant. No amplitude factor is needed, since the data
are already normalized (to check this point, a few calculations have
been done with such a factor, showing that it remains equal to 1). As
above, only the effective vertices $v_1$ and $v_2$ are allowed to vary
with $\phi$.  This means in particular that constant values of
$\tau_0$ and $\tau_1$ are imposed, despite the fact that the $\tau_q$
values of the original equations actually change with compacity
through $S_q$.

\begin{figure}
\rotatebox{270}{\includegraphics*[12.5cm,1.5cm][19.5cm,10cm]{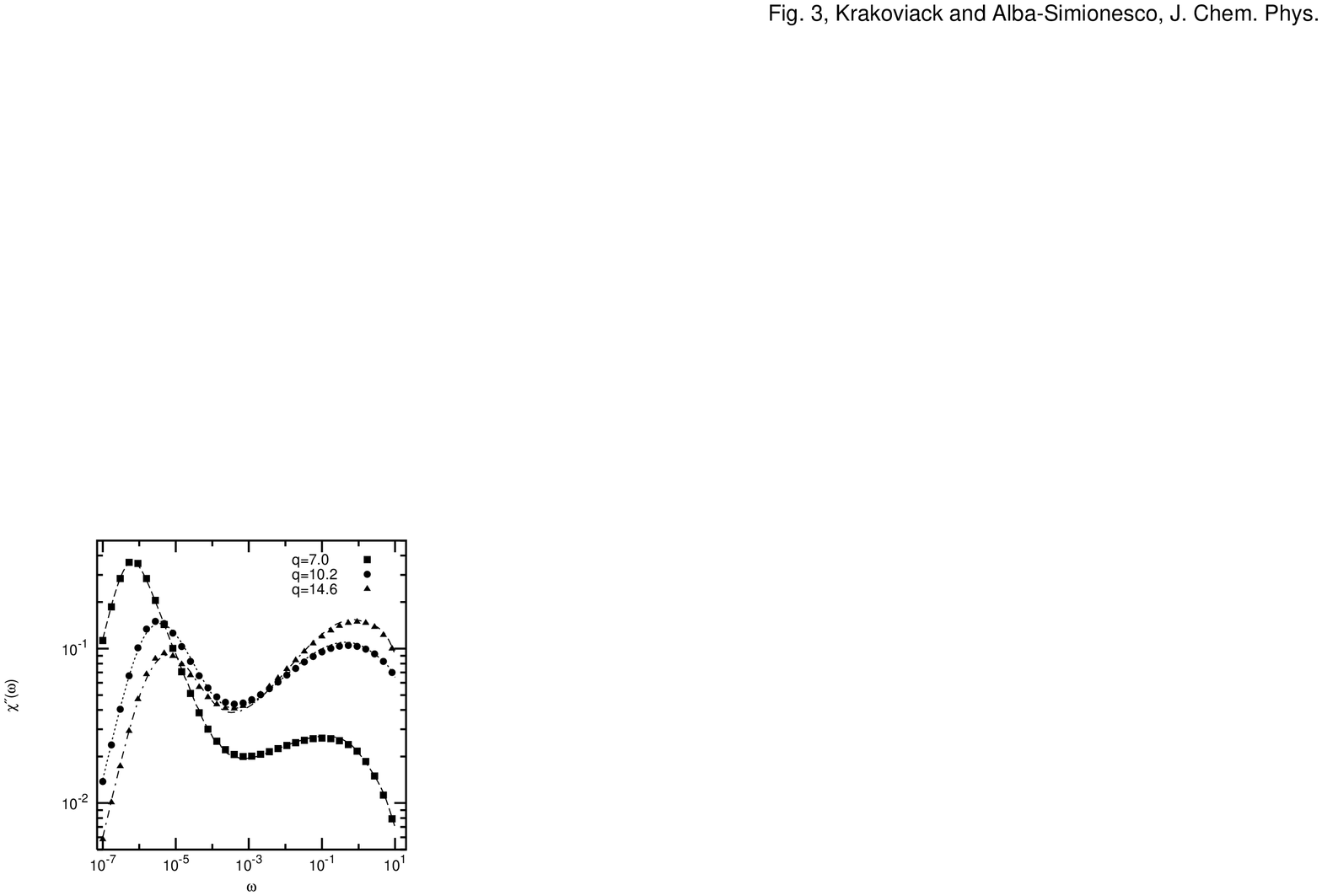}}
\caption{\label{fiths} Susceptibility spectra of a model simple
liquid (symbols) and fits obtained with an F$_{12}$ schematic model
(continuous lines), for $n=8$ and three values of the wave vector. The
size of the symbols has been chosen to roughly model the typical
dispersion of data points obtained from experiments.}
\end{figure}

\begin{figure*}
\rotatebox{270}{\includegraphics*{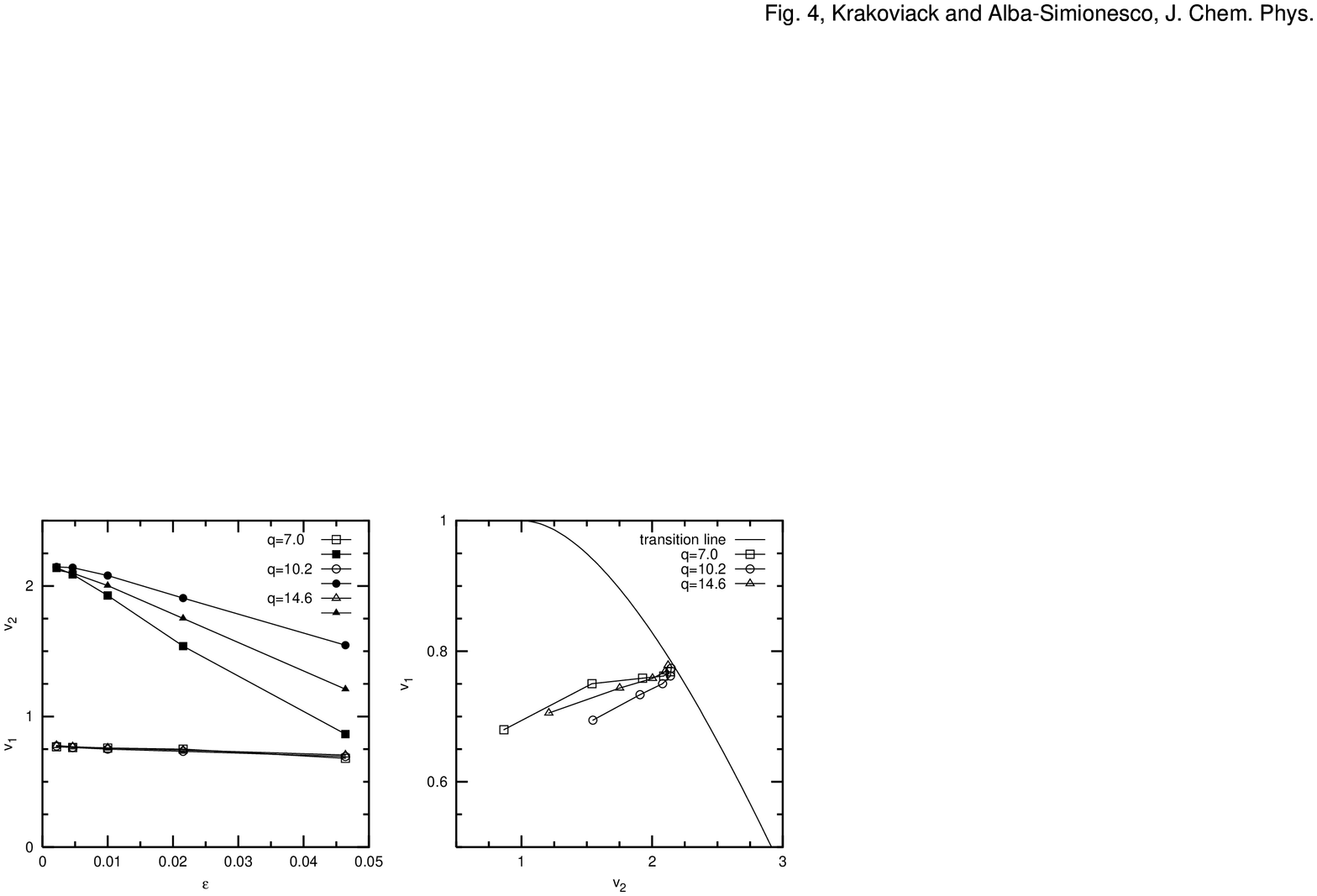}}
\caption{\label{hsvert} Effective vertices obtained from a
schematic analysis with an F$_{12}$ model of the susceptibility
spectra associated to the density fluctuation correlation functions of
a model simple liquid at three wave vector modulus values. Left:
effective vertices as a function of the reduced compacity
$\varepsilon$. Right: trajectories in the control parameter plane
$(v_1,v_2)$.}
\end{figure*}

The quality of the fits is good, comparable to what is obtained when
experimental light-scattering spectra are analyzed, as illustrated in
Fig.~\ref{fiths}. Values obtained for the various adjustable
mode-coupling parameters are reported in Table~\ref{hstab} and in
Fig.~\ref{hsvert}.

\begin{table}
\begin{tabular}{*{7}{c}}
$q$ & \multicolumn{2}{c}{7} & \multicolumn{2}{c}{10.2} &
\multicolumn{2}{c}{14.6} \\ & schem. & theor. & schem. & theor. &
schem. & theor. \\\hline 
$r$ & 20.3 & & 1.8 & & 0.58 \\ 
$\gamma$& 0.102 && 0.329 && 0.503 \\ \hline 
$\phi^c$ & 0.5153& 0.5159 & 0.5155 & 0.5159 & 0.5154 & 0.5159 \\ 
$\lambda$ & 0.678 & 0.735 & 0.676 & 0.735 & 0.679 & 0.735 \\ 
$f_q^c$& 0.846 & 0.849 & 0.414 & 0.419 & 0.268 & 0.269 \\ 
$h_q$ & 0.306 & 0.323 & 0.738 & 0.642 & 0.603 & 0.582 \\ 
$C_\phi$ & 2.48 & 1.54 & 1.36 & 1.54 & 1.81 & 1.54 \\ 
$H_q$ & 0.849 & 0.779 & 1.513 & 1.548 & 1.433 & 1.403
\end{tabular}
\caption{\label{hstab} \small Comparison of the critical parameter
values obtained from a schematic analysis of model spectra and from a
direct study of Eqs. \eqref{hsmodel}.}
\end{table}

As usual, the effective vertices are the important parameters of the
calculation. To each series of curves corresponds a different set of
effective vertices, stressing the fact that the study of a single
observable is not enough to give an unequivocal account of the
dynamics of the system. Here again, this is a consequence of the
non-universality of the $\alpha$ relaxation as described in the MCT
framework. The important observation is that the behavior of the
effective vertices is perfectly regular: $v_1$ and $v_2$ are
approximately linear with $\varepsilon$ (and thus $v_1$ is linear with
$v_2$). Nothing similar to what has been found with experimental data
close to the transition line shows up in the results. This tends to
confirm the assumption that the anomalous behavior found from
experiments and attributed to the avoidance of the ideal glass
transition is not connected to a technical limitation of the schematic
approach, but actually, at a more fundamental level, to the
inadequateness of ideal MCT in a temperature domain starting
noticeably above the extrapolated ideal glass transition point.

We can now determine the critical parameter values by extrapolating
the linear behavior we have found and compare the results with those
obtained directly from the original equations, in order to see if the
procedure we have used previously is meaningful and if the
corresponding results are quantitative. The values are reported in
Table~\ref{hstab}. Of the three parameters whose importance has been
stressed in Section~\ref{sec3}, $f_q^c$ is clearly the one for which
the agreement between the original value and the schematic calculation
is the best. The determination of $H_q$ is less precise but remains
satisfactory. On the contrary, a notable disagreement on the value of
the crucial exponent parameter $\lambda$ is found. The difference we
find is significant and shows unambiguously that the schematic
approach does not allow \emph{a priori} to obtain the exact value of
the exponent parameter, at least when the studied dynamics only covers
a frequency range of the order of four or five decades. From this
point of view, the schematic approach suffers from the same
difficulties as the use of the $\beta$ correlator, which has been
shown not to be able to give the value of $\lambda$ in the same
conditions \cite{scitar99jpcm}. It has however to be stressed that,
even if it is not perfect, the schematic approach turns out to perform
much better than the latter approach. Surprisingly, it is found that
the three series of curves do not provide values scattered around the
exact value $\lambda=0.735$, but instead, the same underestimated
value is consistently found at all three wave numbers. It seems thus
very likely that no improvement would be obtained if the three
observables were analyzed in a unified framework, in the spirit of the
most recent schematic calculations
\cite{rufbeatouecolesmar98jncs,gotvoi00pre}.

The origin of the present problem is strictly connected to the
existence in the analyzed data of corrections to the asymptotic
behavior predicted by the theory and not to a deficiency of the
schematic approach. Indeed, when spectra corresponding to larger
values of $n$ (i.e. corresponding to compacity values closer to the
critical one) are analyzed, a slight change in the behavior of the
effective vertices occurs around $n=10$ and the trajectory described
in the $(v_1,v_2)$ plane bends to indicate the point of the transition
line corresponding to the correct value of $\lambda$. We stress, to
avoid any ambiguity, that this change of behavior has nothing to do
with the one found from experimental data: its amplitude is very small
and it occurs in a domain where the points representative of the
system in the $(v_1,v_2)$ plane are extremely close to the ideal glass
transition line, meaning that the dynamics spreads over more than
eight decades in frequency and corresponding for a real glassformer
like CKN to a regime where the behavior of the effective vertices is
already strongly affected by the existence of activated processes.

\section{Conclusion and perspectives}

In this Paper we have proposed a detailed study of the schematic
mode-coupling approach to experimental data. Our aim was twofold.

First, we wanted to settle the whole approach on firmer theoretical
grounds, through a better understanding of the parameterization it
involves and a clarification of the nature of the results it can
provide. Indeed, the schematic approach involves a certain degree of
arbitrariness in the choice of the models to be used and the
introduction of numerous adjustable parameters, which are never
considered as good things when it comes to testing a theory. Thus in
order to credibilize the whole approach, one has to address these
\textit{a priori} problematic issues and show that they are actually
harmless. Hence this investigation, whose outcome is the formulation
of simple rules that need to be taken into account in order to define
minimal schematic models suitable for use in this context and the
isolation of a few parameters which are particularly relevant for the
description of the dynamics and for which the schematic approach
should provide estimates. Following these criteria, it turns out that
most previous schematic studies
\cite{albkra95jcp,albkrakramigrom96jrs,kraalbkra97jcp,%
fragotmaysin97pre,sinligotfucfracum98jncs,rufbeatouecolesmar98jncs}
are based on well designed schematic models and use parameterizations
which are close to the minimal one, in particular by letting only
effective vertices vary with temperature.

Secondly, our goal was a validation of the approach, through the
investigation of its robustness and of its potential quantitative
character by a variety of numerical tests. The results are rather
satisfactory. Indeed, we have been able to validate the results
obtained in Ref.~\onlinecite{kraalbkra97jcp} and the way they were
analyzed. We have shown that they have a limited, if any, dependency
in the schematic model we use and we have been able to identify and
rule out many potential biases that could have affected our
results. Most properties of the dynamics whose relevance have been
emphasized in the theoretical discussion of the approach appear to be
quantitatively evaluated, with the notable exception of the exponent
parameter $\lambda$, for which the method is not too bad, but still
far from perfect. It has thus to be stressed that the precise and
unambiguous determination of this parameter, which actually plays a
very important role in the theory, remains elusive, by any of the
methods that are available so far: all calculations indeed seem to be
too sensitive to the presence of significant pre-asymptotic
corrections (as is shown here or, in the case of the use of the
$\beta$ correlator, by Ref.~\onlinecite{scitar99jpcm}) and very
probably to the existence of crossover effects due to the non-ideal
behavior of real and simulated glassformers \cite{koband94prl}.

In conclusion, we believe that the present work paves the way to new
applications of the schematic approach. Indeed, the interest of the
method to analyze experimental data has been emphasized in all
previous works and we think its relevance has been further
strengthened here. However, at this level, it was only a kind of
simple phenomenological mode-coupling approach and the main issue was
its ability to consistently describe a set of experimental data with a
simple model. Now, from our results, it seems possible to go one step
further and to use this approach as an effective means to obtain
precise estimates of transition temperatures, non-ergodicity
parameters and critical amplitudes to be used for instance in
quantitative tests of the MCT on simple systems that are amenable to a
full first-principle treatment.

\begin{acknowledgments}
The authors are indebted to Prof.~W.~G\"{o}tze and Dr.~M.~Fuchs for
providing them the codes for some parts of the calculations. We are
also grateful to Prof.~H.Z.~Cummins and Dr.~G.~Li for the use of
their experimental data. Dr.~G.~Tarjus is acknowledged for fruitful
discussions.
\end{acknowledgments}


\begin{thebibliography}{10}
\bibitem{note1} Schematic models have been used earlier to fit
  experimental data \cite{gotsjo87jpc,kribos87prl,fucgothoflat91jpcm},
  but only to give an account of the shape of the $\alpha$ relaxation,
  and not of the evolution with temperature of the full dynamics as in
  the more recent schematic approaches.
\end{thebibliography}
\end{document}